\documentclass[aps,prd,twocolumn,showpacs,amsmath,amssymb,superscriptaddress,groupedaddress]{revtex4-1}

\usepackage{amssymb}
\usepackage{amsfonts}
\usepackage{amsmath,array}
\usepackage{epsfig}
\usepackage{graphicx}
\usepackage{epstopdf} 
\usepackage{dcolumn}
\usepackage{bm}
\usepackage{lineno}
\usepackage{overpic}
\usepackage{multirow}
\usepackage{color}
\usepackage{bm}
\usepackage[utf8]{inputenc}
\usepackage[perpage,symbol]{footmisc}
\usepackage{subfigure}
\usepackage{enumitem}

\usepackage{bm}
\usepackage{rotating}
\usepackage[utf8]{inputenc}

\newcommand{\Br}{\mathcal{B}}

\usepackage[bookmarksnumbered, pdfstartview=FitH,colorlinks,urlcolor=blue, citecolor=blue,linkcolor=blue] {hyperref}

\let\oldequation\equation
\let\oldendequation\endequation

\renewenvironment{equation}
  {\linenomathNonumbers\oldequation}
  {\oldendequation\endlinenomath}

\newcommand{\bes}{BES\uppercase\expandafter{\romannumeral3} }

\begin{document}

\title{\bf \boldmath Improved measurement of the decays $\eta' \to \pi^{+}\pi^{-}\pi^{+(0)}\pi^{-(0)}$ and search for the rare decay $\eta' \to 4\pi^{0}$}

\author{
\begin{small}
  \begin{center}
M.~Ablikim$^{1}$, M.~N.~Achasov$^{4,b}$, P.~Adlarson$^{75}$, X.~C.~Ai$^{80}$, R.~Aliberti$^{35}$, A.~Amoroso$^{74A,74C}$, M.~R.~An$^{39}$, Q.~An$^{71,58}$, Y.~Bai$^{57}$, O.~Bakina$^{36}$
% ...
\\
\vspace{0.2cm}
(BESIII Collaboration)\\
\vspace{0.2cm} {\it
$^{1}$ Institute of High Energy Physics, Beijing 100049, People's Republic of China\\
% ...
$^{4}$ Budker Institute of Nuclear Physics SB RAS (BINP), Novosibirsk 630090, Russia\\
% ...
$^{35}$ Johannes Gutenberg University of Mainz, Johann-Joachim-Becher-Weg 45, D-55099 Mainz, Germany\\
$^{36}$ Joint Institute for Nuclear Research, 141980 Dubna, Moscow region, Russia\\
% ...
$^{39}$ Liaoning Normal University, Dalian 116029, People's Republic of China\\
% ...
$^{57}$ Southeast University, Nanjing 211100, People's Republic of China\\
$^{58}$ State Key Laboratory of Particle Detection and Electronics, Beijing 100049, Hefei 230026, People's Republic of China\\
% ...
$^{71}$ University of Science and Technology of China, Hefei 230026, People's Republic of China\\
% ...
$^{74}$ University of Turin and INFN, (A)University of Turin, I-10125, Turin, Italy; (B)University of Eastern Piedmont, I-15121, Alessandria, Italy; (C)INFN, I-10125, Turin, Italy\\
$^{75}$ Uppsala University, Box 516, SE-75120 Uppsala, Sweden\\
% ...
$^{80}$ Zhengzhou University, Zhengzhou 450001, People's Republic of China\\

\vspace{0.2cm}
% ...
$^{b}$ Also at the Novosibirsk State University, Novosibirsk, 630090, Russia\\
% ...

}

\end{center}
\end{small}
}

\begin{abstract}
  Using a sample of 10 billion $J/{\psi}$ events collected with the BESIII detector,  the decays $\eta' \to \pi^{+}\pi^{-}\pi^{+}\pi^{-}$, $\eta' \to \pi^{+}\pi^{-}\pi^{0}\pi^{0}$ and $\eta' \to 4 \pi^{0}$ are studied via the process  $J/{\psi}\to\gamma\eta'$. The branching fractions of $\eta' \to \pi^{+}\pi^{-}\pi^{+}\pi^{-}$ and $\eta' \to \pi^{+}\pi^{-}\pi^{0}$
$\pi^{0}$ are measured to be $( 8.56 \pm 0.25({\rm stat.}) \pm 0.23({\rm syst.}) ) \times {10^{ - 5}}$ and  $(2.12 \pm 0.12({\rm stat.}) \pm 0.10({\rm syst.})) \times {10^{ - 4}}$, respectively, which are consistent with previous measurements but with improved precision.
 No significant $\eta' \to 4 \pi^{0}$ signal is observed, and the upper limit on the branching fraction of this decay is determined to be less than $1.24 \times {10^{-5}}$ at the $90\%$ confidence level. In addition, an amplitude analysis of $\eta' \to \pi^{+}\pi^{-}\pi^{+}\pi^{-}$ is performed to extract the doubly virtual isovector form factor $\alpha$ for the first time. The measured value of $\alpha=1.22 \pm 0.33({\rm stat.}) \pm 0.04({\rm syst.})$, is in agreement with the prediction of the VMD model.

\end{abstract}

\maketitle

\oddsidemargin  -0.2cm
\evensidemargin -0.2cm

\section{Introduction}

The $\eta^\prime$ meson, interpreted as a flavor singlet state arising from the axial U(1) anomaly, has attracted both theoretical and experimental attention due to its special role in helping to understand low-energy Quantum Chromodynamics (QCD).
Its dominant radiative and hadronic decays have been observed and well measured, however the study of its rare decays is still an open field, and is of value for investigating the symmetry-breaking mechanisms and testing chiral perturbation theory (ChPT). 
In addition, $\eta^\prime$ decays also play an important role in the evaluation of the hadronic
light-by-light contribution to the muon anomalous magnetic moment (see Ref.~\cite{Aoyama:2020,Fang:2021hyq} for details).

 The hadronic decays $\eta^{\prime} \to \pi^+\pi^- \pi^{+(0)}\pi^{-(0)}$, where
the pion pairs are formed by internal conversion of two
intermediate virtual photons, are of special interest.  Within the framework of the Vector Meson Dominance (VMD) model,  one can expect that each virtual photon
converts mainly into a  virtual $\rho$ meson, which decays with 
large probability into two pions, despite the smaller available phase space.  Therefore,  the $\eta^{\prime} \to \pi^+\pi^- \pi^{+(0)}\pi^{-(0)}$  channel can
provide valuable information about the meson-$\gamma^*\gamma^*$ coupling, which is of importance for the calculation of the hadronic contribution to the anomalous magnetic moment of the muon.
However, the information on such decays is scarce. In 2014, the BESIII collaboration reported the first  observation of
$\eta^{\prime} \to \pi^+\pi^- \pi^{+(0)}\pi^{-(0)}$ decays \cite{A1-bes3-eta4pi-lihj}.  The measured branching fractions
are in agreement with theoretical predictions~\cite{GuoFK}, but an amplitude analysis of these decays was not performed due to the limited sample sizes. 

Interest in the very rare decay  $\eta^{\prime} \to 4\pi^0$ stems from the  $S$-wave $C\!P$-violation. As a result,  this decay is highly suppressed and the S-wave $C\!P$-violating effect that contributes to this decay is
at a level of $10^{-23}$~\cite{pich:1991,ottnad:2010}, which is beyond experimental accessibility.  However, within the ChPT and VMD models, the branching fraction may reach a level of $10^{-8}$ through the presence of a $D$-wave contribution~\cite{GuoFK}. Several attempts have been made to search for this decay~\cite{gams4pi, alde:1987, A2-bes3-eta4pi-chang}.   The most stringent upper limit of the branching fraction ${\mathcal{B}}({\eta'}\to 4{\pi}^{0}) < 4.94\times 10^{-5}$ at the 90\% confidence level was set by the BESIII collaboration, from an analysis based on a sample of  $1.31\times 10^{9}$ $J/{\psi}$ events~\cite{A2-bes3-eta4pi-chang}.

BESIII has now collected a data set of $(10087\pm44)\times10^{6}$ ${J/{\psi}}$ events~\cite{yanghx}, which allows for improved analyses of $\eta'$ decaying into   four pions and also the extraction of the
timelike double virtual transition form factor (TFF) information of the $\eta^\prime$ meson. In this paper, we present studies of ${\eta'}\to{\pi}^{+}{\pi}^{-}\pi^{+(0)}\pi^{-(0)}$ and ${\eta'}\to 4{\pi}^{0}$ via the process $J/{\psi} \to \gamma {\eta'}$. In addition, an amplitude analysis of $\eta' \to \pi^{+} \pi^{-} \pi^{+} \pi^{-}$ is performed to extract the doubly virtual time-like form factor of $\eta'$ within the framework of the combination of the ChPT and VMD models~\cite{GuoFK}.

\section{Detector and Monte Carlo samples}
The BESIII detector \cite{dector} records symmetric $e^{+} e^{-}$ collisions provided by the BEPCII storage ring \cite{ring} in the center-of-mass energy range from 2.0 to $4.95~{\rm GeV}$, with a peak luminosity of 1.1 $\times 10^{33} {\rm cm}^{-2} s^{-1}$ achieved at $\sqrt{s} = 3.77$ GeV. The BESIII detector has collected large data samples in this energy region \cite{region}. The cylindrical core of the BESIII detector covers 93\% of the full solid angle and comprises a helium-based multilayer drift chamber (MDC), a plastic scintillator time-of-flight system (TOF), and a CsI (TI) electromagnetic calorimeter (EMC), which are all enclosed in a superconducting solenoidal magnet providing a 1.0~T magnetic field. The solenoid is supported by an octagonal flux-return yoke with resistive plate counter muon identification modules interleaved with steel. The charged-particle momentum resolution at 1~GeV/c is 0.5\%, and the ${\rm d}E/{\rm d}x$ resolution is 6\% for electrons from Bhabha scattering. The EMC measures photon energies with a resolution of 2.5\% (5\%) at~1 GeV in the barrel (end-cap) region. The time resolution in the TOF barrel region is 68~ps, while that in the end-cap region is 110~ps. The end-cap TOF system was upgraded in 2015 using multi-gap resistive plate chamber technology, providing a time resolution of 60 ps.

Simulated data samples produced with a {\sc geant4}-based \cite{geant4} Monte Carlo (MC) package, which
includes the geometric description of the BESIII detector and the
detector response, are used to determine the detection efficiencies and
estimate backgrounds.
The simulation models the beam energy spread and initial-state radiation in the $e^{+} e^{-}$ annihilations with the generator {\sc kkmc} \cite{kkmc}. The inclusive MC sample of 10 billion $J/\psi$ events consists of the production of $J/\psi$ resonance, and the continuum processes incorporation in {\sc kkmc} \cite{kkmc}.  
All particle decays are modeled with {\sc evtgen} \cite{evtgen} using branching fractions either taken from the Particle Data Group (PDG) \cite{pdg}, when available, or otherwise estimated with {\sc lundcharm} \cite{lund}.
Final-state radiation from charged final-state particles is incorporated using the {\sc photos} package \cite{photos}. Specific generators are employed for three signal channels of $\eta' \to \pi^{+} \pi^{-} \pi^{+(0)} \pi^{-(0)}$ and $\eta' \to 4 \pi^{0}$, which are based on the ChPT and VMD models~\cite{GuoFK}.

\section{Event selection and Background analysis}
Each charged track, reconstructed in the MDC, is required to originate from a region within 10~cm of the interaction point (IP) along the beam direction and 1~cm in the plane perpendicular to the beam. The polar angle $\theta$ of the tracks must be within the fiducial volume of the MDC $\left| {\cos \theta } \right| < 0.93$. Photon candidates are identified using showers in the EMC. The deposited energy of each shower must be more than 25~MeV in the barrel region $\left( {\left| {\cos \theta } \right| < 0.8} \right)$ and more than 50~MeV in the end-cap region $\left( {0.86<\left| {\cos \theta } \right| < 0.92} \right)$. The difference between the EMC time and the event start time is required to be within ($0,700$)~ns  {for the decay modes $\eta' \to \pi^{+} \pi^{-} \pi^{+(0)} \pi^{-(0)}$,
while the timing of the shower with respect to the most energetic photon must lie within ($-500,500$)~ns for the decay $\eta' \to 4 \pi^{0}$ to suppress electronic noise and showers unrelated to the event}.

\begin{figure}[htp]
  \begin{center}
    \subfigure{
    \centering
      \label{m4pi}
      \includegraphics[width=0.49\textwidth]{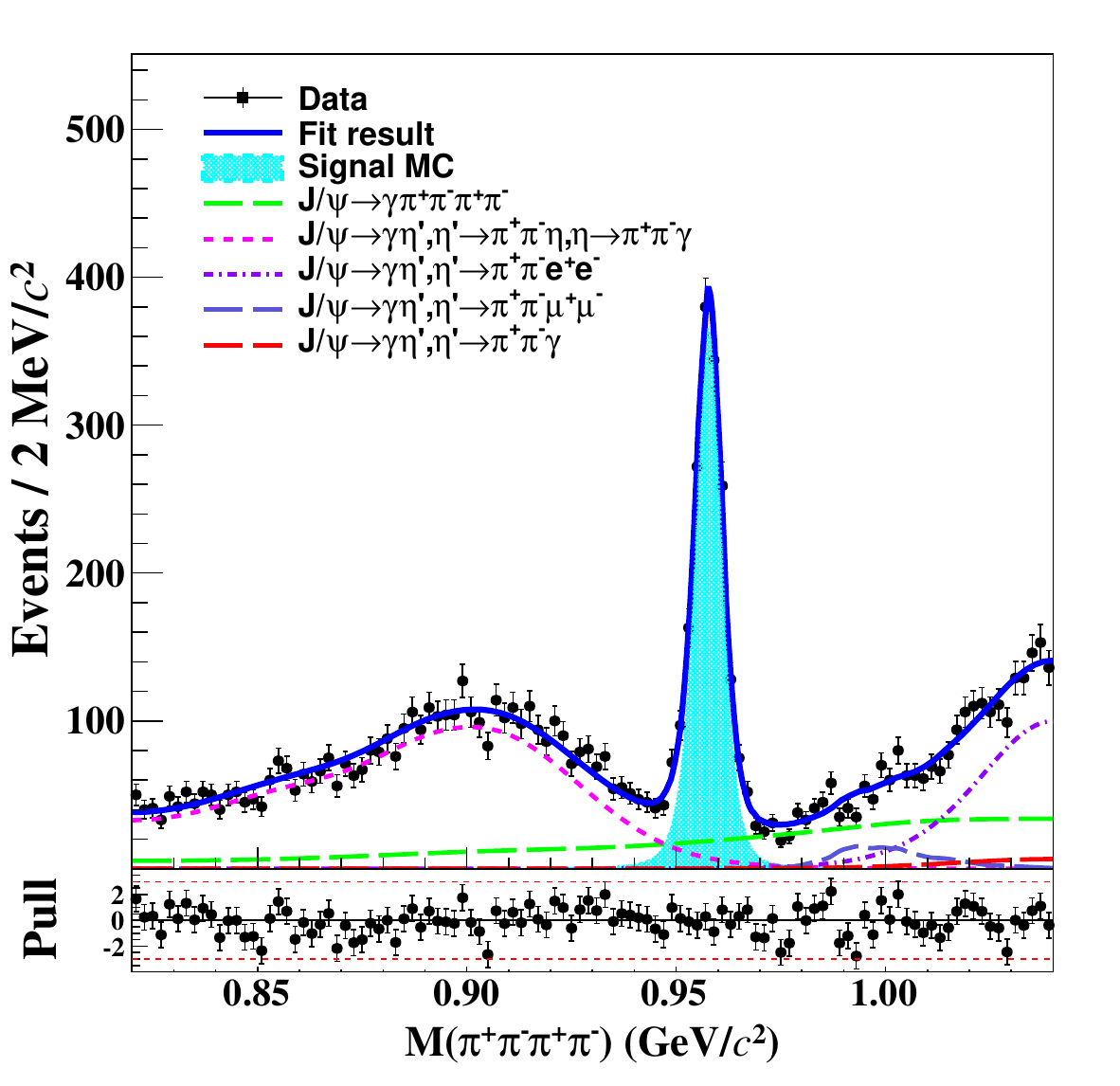}
      \put(-40,215){\bf~(a)}
    }
    \subfigure{
    \centering
      \label{m4pi_2pi0}
      \includegraphics[width=0.49\textwidth]{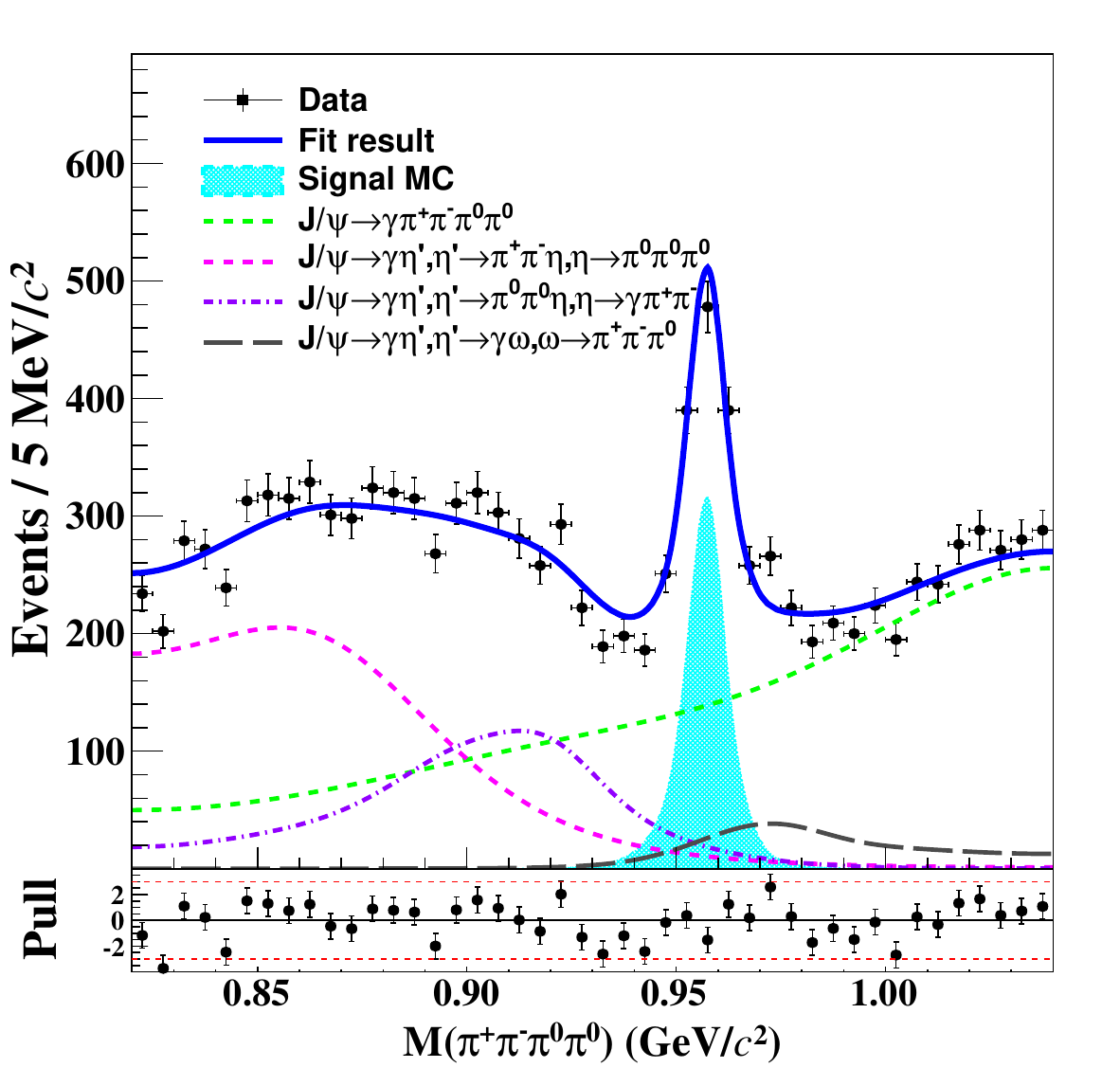}
      \put(-40,215){\bf~(b)}
    }
    \caption{The fit results of  the mass spectra of $\pi^{+} \pi^{-} \pi^{+} \pi^{-} $ (a) and $\pi^{+} \pi^{-} \pi^{0} \pi^{0} $(b). 
}
    \label{fig:m_m4pifit}
  \end{center}
\end{figure}

For $J/{\psi} \to \gamma \eta'$ with $ \eta' \to \pi^{+} \pi^{-} \pi^{+} \pi^{-}$ (Mode I), candidate events are required to have four charged tracks with net zero charge and at least one photon. The candidate events are required to successfully pass a vertex fit to the interaction point (IP). Then a four-constraint (4C) kinematic fit to the initial $e^+e^-$ four momentum is performed under the $\gamma \pi^{+} \pi^{-} \pi^{+} \pi^{-}$
hypothesis. In events with more than one photon, all combinations are considered under this hypothesis and the combination with the smallest $\chi^{2}_{\rm 4C}$ is retained for further analysis. 
To reject possible background events with one more photon,  we further require that the probability of the 4C fit for the $\gamma \pi^{+} \pi^{-} \pi^{+} \pi^{-}$ assignment is greater than that for the $\gamma \gamma \pi^{+} \pi^{-} \pi^{+} \pi^{-}$ hypothesis.
With the requirement of  $\chi^{2}_{\rm 4C}< 35$, the $\pi^+\pi^-\pi^+\pi^-$ mass spectrum is displayed in Fig.~\ref{m4pi}, where a clear $\eta^\prime$ peak is observed. To ensure that the $\eta'$ peak originates from the signal process rather than background, the same analysis selection is performed with the inclusive MC sample of 10 billion $J/{\psi}$ events.
Detailed event-type analysis~\cite{topology} over the surviving MC candidates indicates that the broad enhancement below the $\eta'$ peak is from the background channel $\eta' \to \pi^{+} \pi^{-} \eta$ with $\eta \to \gamma \pi^{+} \pi^{-}$ and the region around  $0.98$ GeV/${c^2}$ receives contamination from $\eta' \to \pi^{+} \pi^{-} \mu^{+} \mu^{-}$ decays. The backgrounds in the mass region above 1 GeV/${c^2}$ are mainly from $\eta' \to \pi^{+} \pi^{-} e^{+} e^{-}$ and $\eta' \to \gamma \pi^{+} \pi^{-}$ events. The remaining  non-resonant background events are from $J/{\psi} \to \gamma \pi^{+} \pi^{-} \pi^{+} \pi^{-}$ events. However, none of these background sources peak close to the $\eta^\prime$ signal in the $\pi^{+} \pi^{-} \pi^{+} \pi^{-}$ mass spectrum.

\begin{figure}[htp]
  \begin{center}
    \subfigure{
      \label{veto_eta}
      \includegraphics[width=0.49\textwidth]{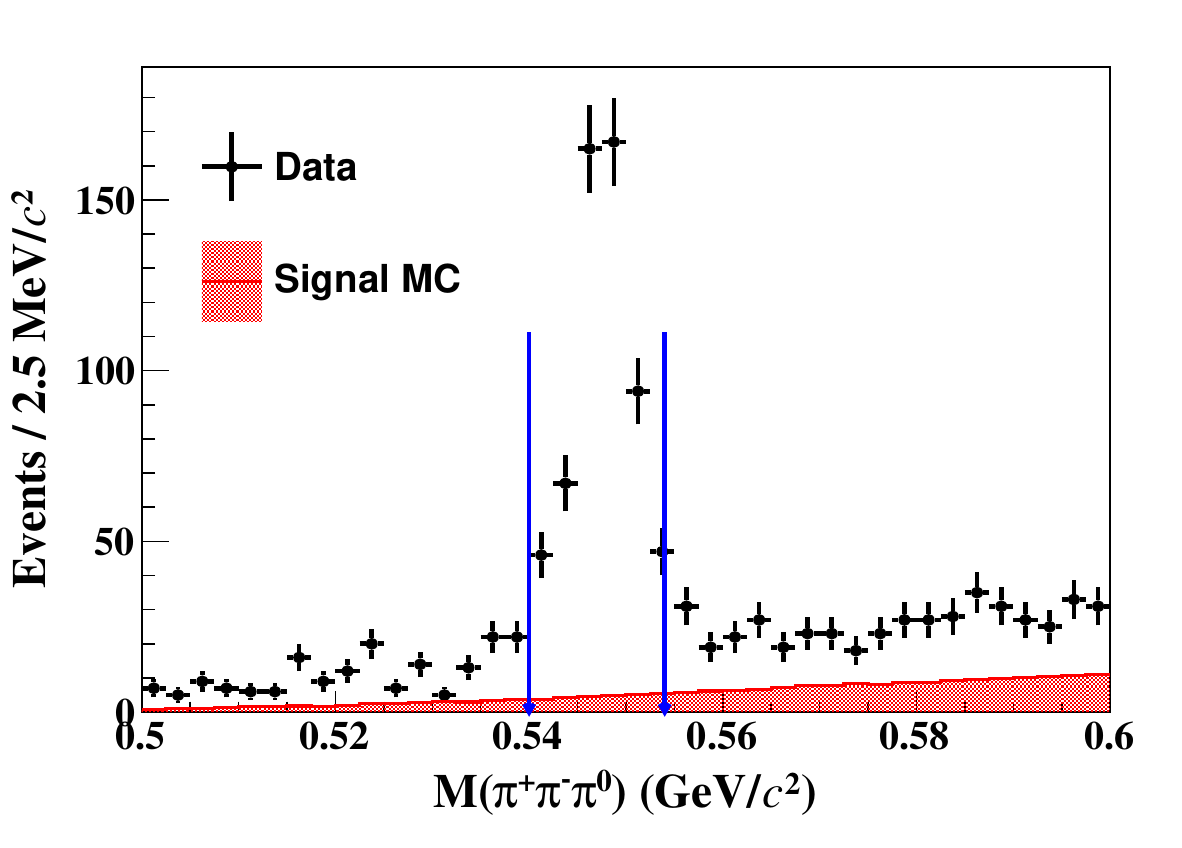}
    \put(-40,150){\bf~(a)}
    }
    \subfigure{
      \label{veto_omega}
      \includegraphics[width=0.49\textwidth]{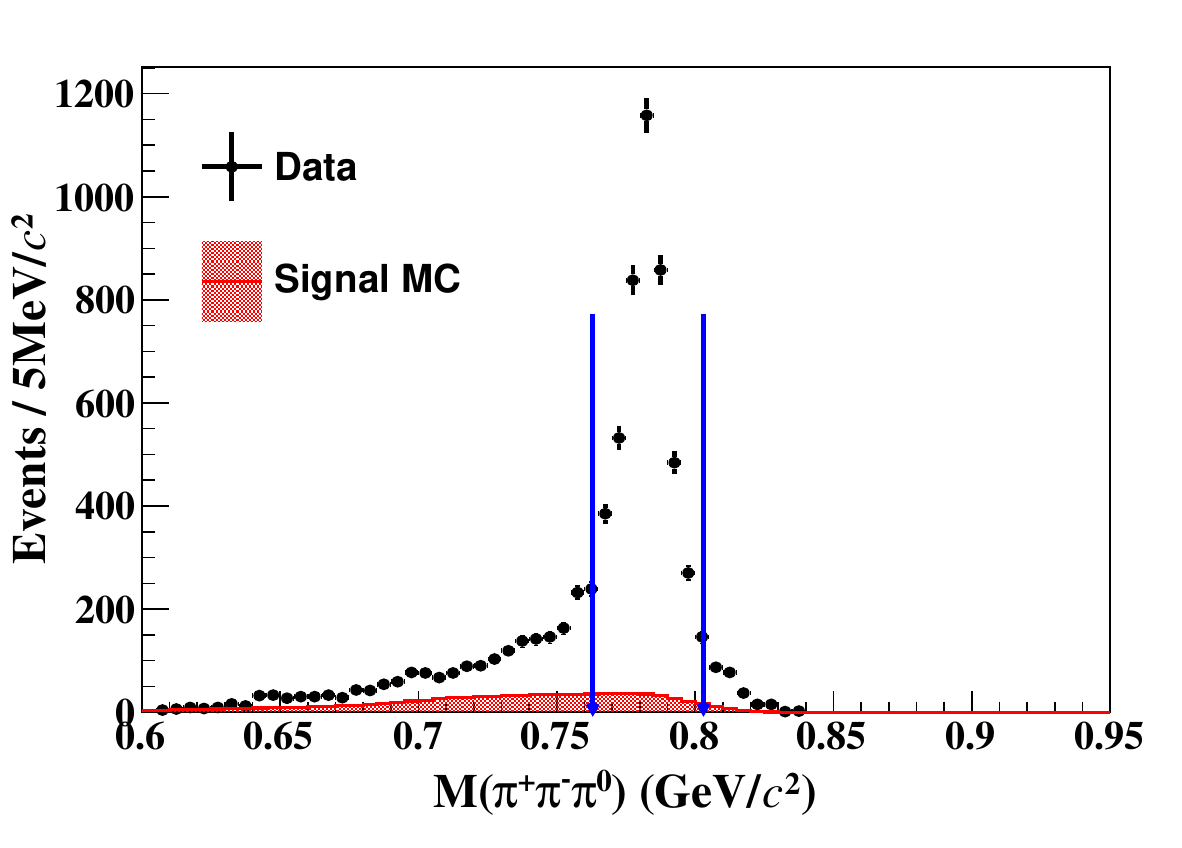}
    \put(-40,150){\bf~(b)}
    }
    \caption{The distributions of the $\pi^+\pi^-\pi^0$ invariant masses closest to the known masses of (a) $\eta$ and (b) $\omega$.
The blue arrows mark the interval for rejecting each resonance.
The signal MC is displayed with arbitrary normalization. 
}
    \label{fig:veto_eta_omg}
  \end{center}
\end{figure}

For $J/{\psi} \to \gamma \eta'$ with $ \eta' \to \pi^{+} \pi^{-} \pi^{0} \pi^{0}$ (Mode II), candidate events must have two charged tracks with zero net charge and at least five good photons.
The charged tracks also need to successfully pass the IP vertex fit.
To reconstruct the $\pi^{0}$ candidate, a one-constraint (1C) kinematic fit is performed on each photon pair with the invariant mass constrained to the known $\pi^{0}$ mass,
and ${\chi}^2_{\rm 1C}(\gamma \gamma) <50$ is required.
Then a six-constraint (6C) kinematic fit to the initial $e^+e^-$ four momentum and the nominal $\pi^0$ masses for two $\gamma\gamma$ pairs is performed under the hypothesis of $J/{\psi} \to \gamma \pi^{+} \pi^{-} \pi^{0} \pi^{0}$ in which the two $\pi^0$ masses are constrained. For events with more than two $\pi^{0}$ candidates, the combination with the smallest ${\chi}^2_{\rm 6C}$ is retained, and ${\chi}^2_{\rm 6C} < 35$ is required to exclude events kinematically incompatible with the signal hypothesis. To reject backgrounds with six photons in the final state, a 6C kinematic fit under the hypothesis of $J/{\psi} \to \gamma \gamma \pi^{+} \pi^{-} \pi^{0} \pi^{0}$ is also made and the ${\chi}^2_{\rm 6C}$ is required to be greater than that of the signal hypothesis $\gamma \pi^{+} \pi^{-} \pi^{0} \pi^{0}$.
Figures~\ref{veto_eta} and~\ref{veto_omega} show the invariant-mass spectra of the $\pi^{+} \pi^{-} \pi^{0}$ combination passing this selection and lying closest to the known $\eta$ ($\omega$) mass (denoted as $m_{\eta}$/$m_{\omega}$),
 respectively, where the $\eta$ and $\omega$ peaks are evident. To reject  background events with $\eta$ ($\omega$) in the final states, the combination closest to $m_{\eta}$ ($m_{\omega}$) is required to satisfy $|M(\pi^{+} \pi^{-}\pi^{0})-m_{\eta}| > 0.007$~GeV/$c^2$ ($|M(\pi^{+} \pi^{-}\pi^{0})-m_{\omega}| > 0.02~$GeV/$c^2$).

Figure~\ref{m4pi_2pi0} shows the  $\pi^{+} \pi^{-} \pi^{0} \pi^{0}$ invariant mass for events satisfying the above requirements, in which a clean $\eta^\prime$ peak is
evident. The same selection is performed on the inclusive MC sample of 10 billion $J/{\psi}$ events to investigate possible sources of contamination.  The dominant backgrounds are found to arise from $\eta' \to \pi^{+} \pi^{-} \eta, \eta \to \pi^{0} \pi^{0} \pi^{0}$,  $\eta' \to \pi^{0} \pi^{0} \eta, \eta \to \gamma \pi^{+} \pi^{-}$,  $\eta' \to \gamma \omega, \omega \to \pi^{+} \pi^{-} \pi^{0}$ and  non-resonant $J/{\psi} \to \gamma \pi^{+} \pi^{-} \pi^{0} \pi^{0}$ events. None of these background channels peak in the $\eta'$ signal region.

When selecting $J/{\psi} \to \gamma \eta'$ with $ \eta' \to 4 \pi^{0}$ (Mode III) events, a one-constraint (1C) kinematic fit is performed on the $\pi^{0}$ candidates reconstructed from photon pairs with the invariant mass of the two photons constrained to the $\pi^{0}$ mass, with the requirement ${\chi}^2_{\rm 1C}(\gamma \gamma) <10$.
A further eight-constraint (8C) kinematic fit to the initial $e^+e^-$ four momentum and the nominal $\pi^0$ masses for four $\gamma\gamma$ pairs is performed to the $\gamma \pi^{0} \pi^{0} \pi^{0} \pi^{0}$ hypothesis by enforcing energy-momentum conservation and constraining the invariant mass of each of the four photon pairs to the known $\pi^{0}$ mass.
If more than one candidate combination is found, that one with the smallest ${\chi^{2}_{\rm 8C}}$ is retained. Candidate events with ${\chi^{2}_{\rm 8C}}>40$ are rejected to exclude events that are kinematically incompatible with the signal hypothesis.
An additional requirement of $\left| M(\pi^{0} \pi^{0} \pi^{0}) - m_{\eta}\right| > 0.02 $ GeV/${c^2}$ is used to remove the $\eta$ background events, where $M(\pi^{0} \pi^{0} \pi^{0})$ is the combination of the $\pi^{0} \pi^{0} \pi^{0}$ invariant mass closest to the known mass of the $\eta$ meson.

To reject background with 10 or 11 photons in the final states, the $\chi^2_{\rm 4C}$ of the 4C kinematic fit to signal hypothesis
is required to be less than those of the 4C kinematic fits to both the 10 and 11 photon hypotheses. To investigate possible sources of contamination, we apply these selections to the inclusive MC sample of 10 billion $J/{\psi}$ events and find that the only significant backgrounds come from  $\eta' \to \pi^{0} \pi^{0} \eta, \eta \to \pi^{0} \pi^{0} \pi^{0}$ and non-resonant $J/{\psi} \to \gamma \pi^{0} \pi^{0} \pi^{0} \pi^{0}$ events.

\section{Measurement of  $\mathcal{B}(\eta' \to \pi^{+} \pi^{-} \pi^{+(0)} \pi^{-(0)})$}
The signal yields of the exclusive channels are obtained by performing unbinned maximum likelihood fits to the mass spectra of the selected $\eta' \to \pi^{+} \pi^{-} \pi^{+} \pi^{-}$ and $\eta' \to \pi^{+} \pi^{-} \pi^{0} \pi^{0}$ candidates, respectively. In the fits, the signal components
are modeled by the MC-simulated shape convolved with a Gaussian function to account for the difference in the mass resolution between data and MC simulation.
 The background components considered are subdivided
into two classes: (i) the shapes of those background events that
contribute to a structure in $M(\pi^+\pi^-\pi^+\pi^-)$  (e.g.
$\eta^\prime\rightarrow\pi^+\pi^-\eta$ with
$\eta\rightarrow\gamma\pi^+\pi^-$ and
$\eta^\prime\rightarrow\pi^+\pi^- e^+e^-$) or
$M(\pi^+\pi^-\pi^0\pi^0)$ ( e.g.
$\eta^{\prime}\to\pi^{+}\pi^{-}\eta$ with
$\eta\to\pi^{0}\pi^{0}\pi^{0}$ and
$\eta^{\prime}\to\pi^{0}\pi^{0}\eta$ with
$\eta\to\gamma\pi^{+}\pi^{-}$, as well as
$\eta^{\prime}\to\gamma\omega$ with
$\omega\to\pi^{+}\pi^{-}\pi^{0}$) are taken from the dedicated MC
simulations;  (ii) $J/\psi\rightarrow\gamma\pi^+\pi^-\pi^+\pi^-$ and 
$J/\psi\rightarrow\gamma\pi^+\pi^-\pi^0\pi^0$ non-resonant decays are
described with the MC-simulated shape in uniform phase-space model. The magnitudes of the different components are
left free
in the fit to account for the
uncertainties of the branching fractions of
$J/\psi\rightarrow\gamma\eta^\prime$ and other intermediate decays
(e.g. $\eta^\prime\rightarrow\pi^+\pi^-\eta$,
$\eta^\prime\rightarrow\pi^0\pi^0\eta$, and
$\eta\rightarrow\gamma\pi^+\pi^-$).

The fits, as shown in Fig.~\ref{fig:m_m4pifit}, result in $1650\pm48$ events for $\eta' \to \pi^{+} \pi^{-} \pi^{+} \pi^{-}$ and $865\pm49$ events for $\eta' \to \pi^{+} \pi^{-} \pi^{0} \pi^{0}$. The detection efficiencies are obtained from the MC simulation in which events are generated with the ChPT and VMD models~\cite{GuoFK}. The measured branching fractions of these two decay modes, as summarized in Table~\ref{list_summary}, are consistent with previous results~\cite{A1-bes3-eta4pi-lihj} obtained from a data set of 1.3 billion $J/\psi$ events, which is a subset of the sample used in the current analysis.

\begin {table*}[htbp]
\begin{center}
    {\caption {The fitted signal yields (or 90\% confidence-level upper limit), the detection efficiencies and the measured branching fractions (or 90\% confidence-level upper limit).  Also shown are the previous BESIII results.
    }
    \label{list_summary}}
    \begin {tabular}{l  c c c c}\hline\hline
Mode             &       $N$               &   $\varepsilon~(\%)$             &    $\Br$($\eta' \to X$)    &    Previous BESIII result~\cite{A1-bes3-eta4pi-lihj,A2-bes3-eta4pi-chang} \\   \hline
	 $\eta' \to \pi^{+}\pi^{-}\pi^{+} \pi^{-}$     &   1650$\pm 48$        & 36.4        &  $\left( { 8.56 \pm 0.25({\rm stat.})\pm 0.23({\rm syst.}) }  \right) \times {10^{ - 5}}$    &  $\left( { 8.53 \pm 0.69({\rm stat.})\pm 0.64({\rm syst.}) }  \right) \times {10^{ - 5}}$\\
    	
     $\eta' \to \pi^{+}\pi^{-}\pi^{0} \pi^{0}$     &   865$\pm 49$        & 7.8        &  $\left( { 2.12 \pm 0.12({\rm stat.}) \pm 0.10({\rm syst.}) }  \right) \times {10^{ - 4}}$    &  $\left( { 1.82 \pm 0.35({\rm stat.}) \pm 0.18({\rm syst.}) }  \right) \times {10^{ - 4}}$\\

	 $\eta' \to \pi^{0}\pi^{0}\pi^{0} \pi^{0}$     &  $ < 10 $       & 1.6        &  $< 1.24   \times {10^{ - 5}}$    &  $< 4.94   \times {10^{ - 5}}$\\

\hline
\hline
\end{tabular}

\end{center}
\end{table*}

\section{Amplitude analysis of $\eta' \to \pi^{+} \pi^{-} \pi^{+} \pi^{-}$}
To investigate the doubly virtual isovector contribution,  an amplitude analysis of $\eta' \to \pi^{+} \pi^{-} \pi^{+(0)} \pi^{-(0)}$ is necessary. Given the limited sample size and high background level for 
$\eta' \to \pi^{+} \pi^{-} \pi^{0} \pi^{0}$ decays, we only perform an amplitude analysis of  $\eta' \to \pi^{+}(p_1) \pi^{-}(p_2) \pi^{+}(p_3) \pi^{-}(p_4)$. The decay amplitude is constructed with a combination of the ChPT and VMD models~\cite{GuoFK},
as described by
\begin{eqnarray}
  \begin{aligned}
    &\mathcal{A}(\eta' \to \pi^{+} \pi^{-} \pi^{+} \pi^{-}) = \epsilon_{\mu\nu\alpha\beta} p_{1}^{\mu} p_{2}^{\nu} p_{3}^{\alpha} p_{4}^{\beta}\\ 
    & \times\Biggl\{\left[\frac{s_{12}}{D_{\rho}(s_{12})} + \frac{s_{34}}{D_{\rho}(s_{34})} - \frac{s_{14}}{D_{\rho}(s_{14})} - \frac{s_{23}}{D_{\rho}(s_{23})}\right]  \\
    &+ \alpha \left[\frac{M^2_{\rho}(s_{12} + s_{34})}{D_{\rho} (s_{12})  D_{\rho} (s_{34})} - \frac{M^2_{\rho}(s_{14} + s_{23})}{D_{\rho} (s_{14})  D_{\rho} (s_{23})}\right]\Biggr\}
    \label{eqfit_amp}
  \end{aligned}
\end{eqnarray}
where
\begin{align}
D_{\rho}(s)=M^{2}_{\rho}-s-i M_{\rho}\Gamma_{\rho}(s),\\
\Gamma_{\rho}(s)=\frac{M_{\rho}}{\sqrt{s}}(\frac{s-4 M^{2}_{\pi}}{M^{2}_{\rho}-4 M^{2}_{\pi}})^{3/2} \Gamma_{\rho},
\end{align}
is the inverse $\rho$ propagator ($M_{\rho}$ and $\Gamma_{\rho}$ are the mass and width of $\rho$ meson), and $s_{12}=(p_1 + p_2)^2$ and $s_{34}=(p_3 + p_4)^2$ are the invariant-masses squared of two independent $\pi^+\pi^-$ pairs; 
$\alpha$ corresponds to $\frac{c_3}{c_1-c_2}$ in Ref.~\cite{GuoFK} since the fit is only sensitive to this ratio rather than the individual parameters. 

The free parameters of the probability-density function (PDF) to observe the $i$-th event characterized by the measured four-momenta $\xi_i$ of the particles in the final state is 
\begin{eqnarray}
  \begin{aligned}
    \mathcal{P}\left(\xi_i\right) = \frac{ \left|\mathcal{A}\left(\xi_i\right)\right|^2 \epsilon\left(\xi_i\right) }{\int \left|\mathcal{A}\left(\xi\right)\right|^2 \epsilon\left(\xi\right) d\xi},
    \label{eqfit_pdf}
  \end{aligned}
\end{eqnarray}
where $\mathcal{A}$ is the amplitude as shown in Eq.~\ref{eqfit_amp}, and $\epsilon\left(\xi_i\right)$ is the detection efficiency. The parameters are optimized with an unbinned maximum likelihood fit using the sample of selected
$\eta' \to \pi^{+} \pi^{-} \pi^{+} \pi^{-}$ events, a total of 2047 events found in the $\eta^\prime$ mass region of $0.94<M(\pi^+\pi^-\pi^+\pi^-) <0.97$ GeV/c$^2$.
The fit minimizes the negative log-likelihood value
\begin{eqnarray}
  \begin{aligned}
    &-\ln{\mathcal {L}} = -\omega'\Biggl[\sum_{i=1}^{N_{data}}\ln\mathcal{P}\left(\xi_i\right) \\
    & -\omega_{bkg1}\sum_{j=1}^{N_{bkg1}}\ln \mathcal{P}\left(\xi_j\right)-\omega_{bkg2}\sum_{k=1}^{N_{bkg2}}\ln \mathcal{P}\left(\xi_k\right)\Biggl],
    \label{eqfit_likelihood}
  \end{aligned}
\end{eqnarray}
where $\mathcal{P}$ is the PDF, $i$, $j$ and $k$ run over all accepted data, the continuous background $J/\psi \to \gamma \pi^{+} \pi^{-} \pi^{+} \pi^{-}$ and the peaking background $J/{\psi} \to \gamma \eta', \eta' \to \pi^{+} \pi^{-}\eta, \eta \to \gamma \pi^{+} \pi^{-}$ events, respectively, and their corresponding number of events are denoted by $N_{data}$, $N_{bkg1}$ and $N_{bkg2}$, with the backgrounds modeled by MC simulation. Here, $\omega_{bkg1}=\frac{N'_{bkg1}}{N_{bkg1}}$ and $\omega_{bkg2}=\frac{N'_{bkg2}}{N_{bkg2}}$ are the weights of the backgrounds, where $N'_{bkg1}$ and $N'_{bkg2}$ are their contributions in the signal region according to the above fit results as displayed in Fig.~\ref{m4pi}. To
obtain an unbiased uncertainty estimation, the normalization factor derived from Ref.~\cite{Langenbruch} is considered, described as
\begin{eqnarray}
  \begin{aligned}
    \omega'=\frac{N_{data}-N_{bkg1}\omega_{bkg1}-N_{bkg2}\omega_{bkg2}}{N_{data}+N_{bkg1}\omega^{2}_{bkg1}+N_{bkg2}\omega^{2}_{bkg2}}
    \label{eqfit_alpha}
  \end{aligned}
\end{eqnarray}

\begin{figure*}[htbp]
\begin{center}
\includegraphics[width=0.25\textwidth]{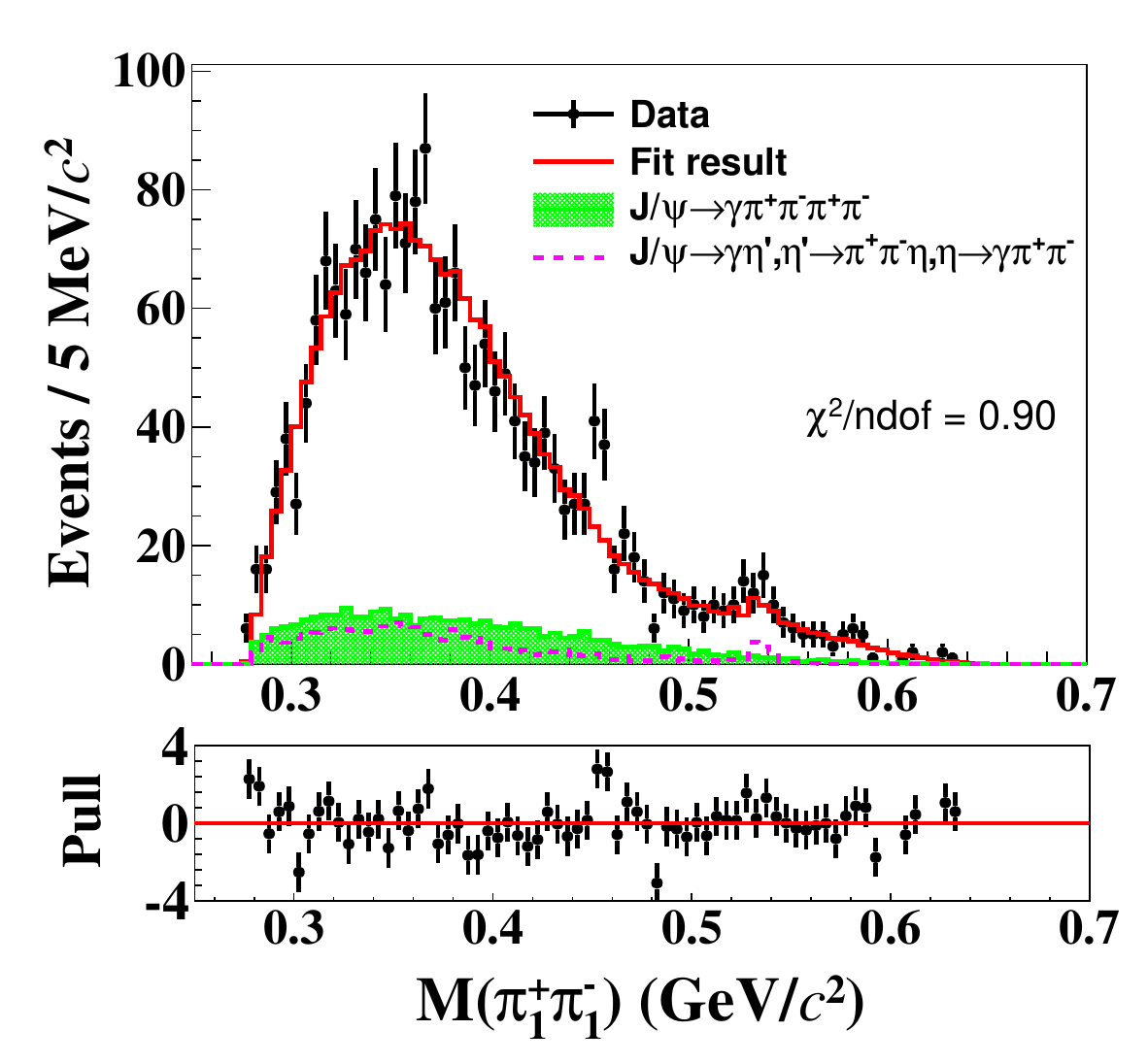}\put(-30,99){\bf~(a)}
\includegraphics[width=0.25\textwidth]{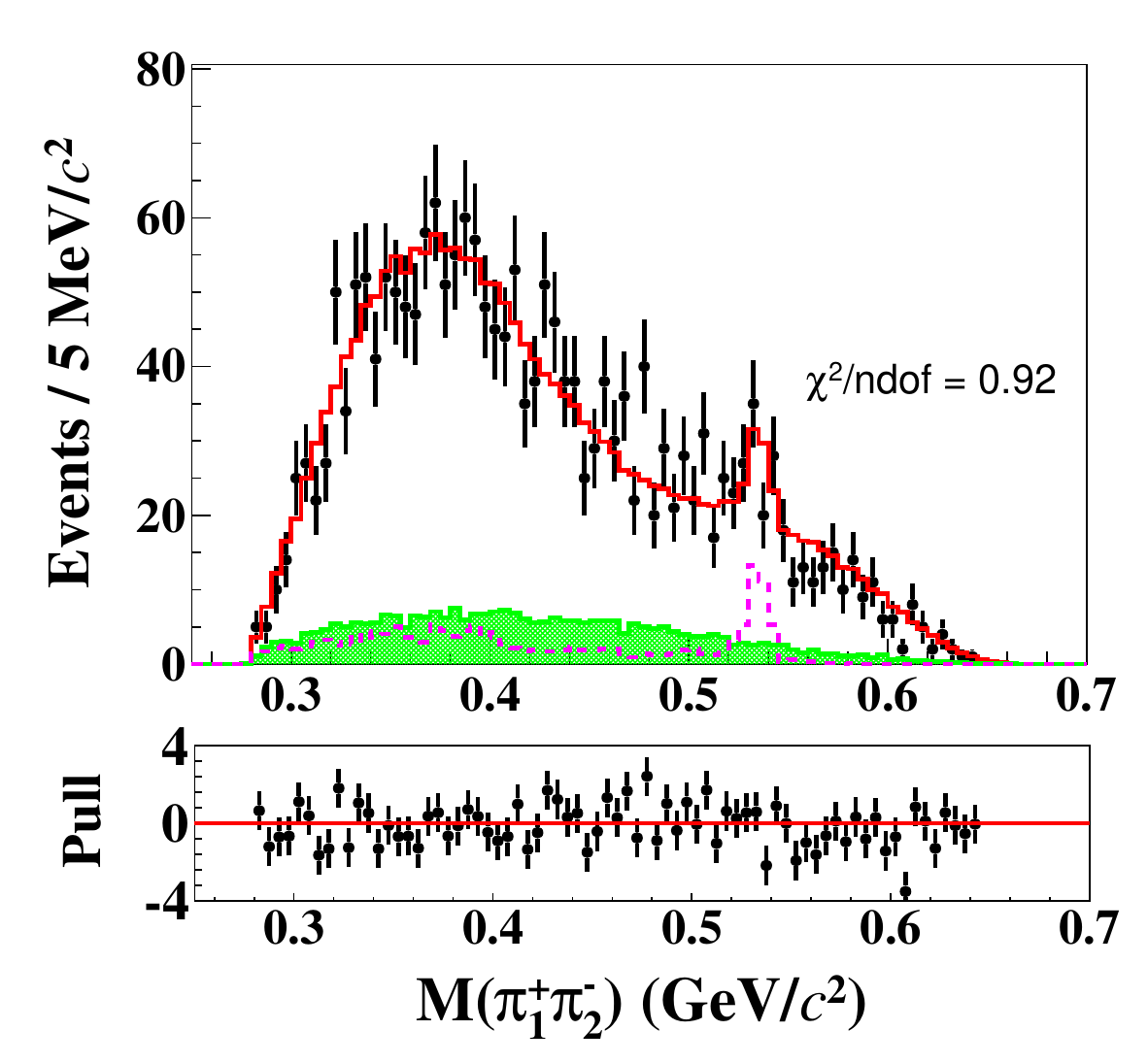}\put(-30,99){\bf~(b)}
\includegraphics[width=0.25\textwidth]{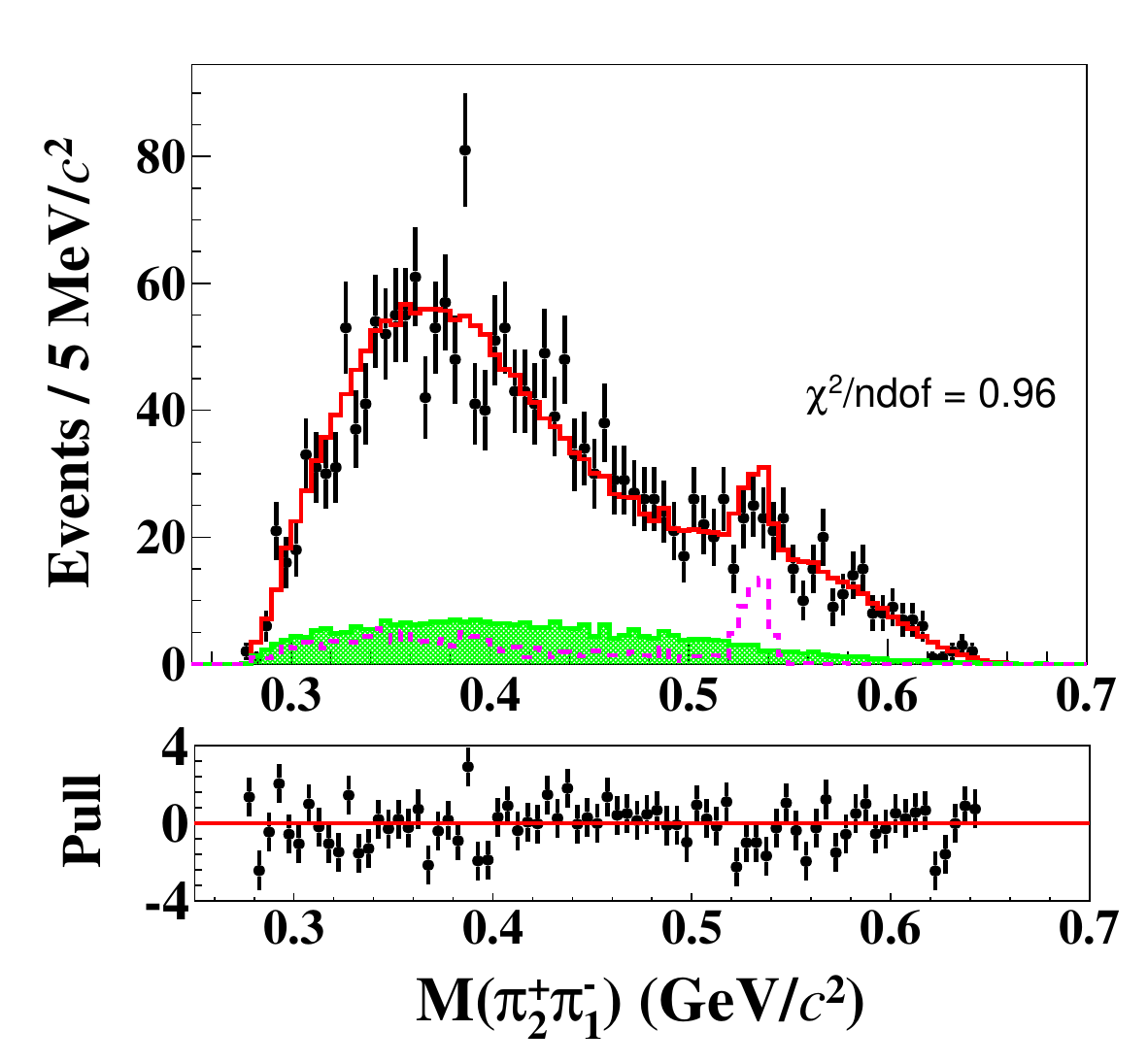}\put(-30,99){\bf~(c)}
\includegraphics[width=0.25\textwidth]{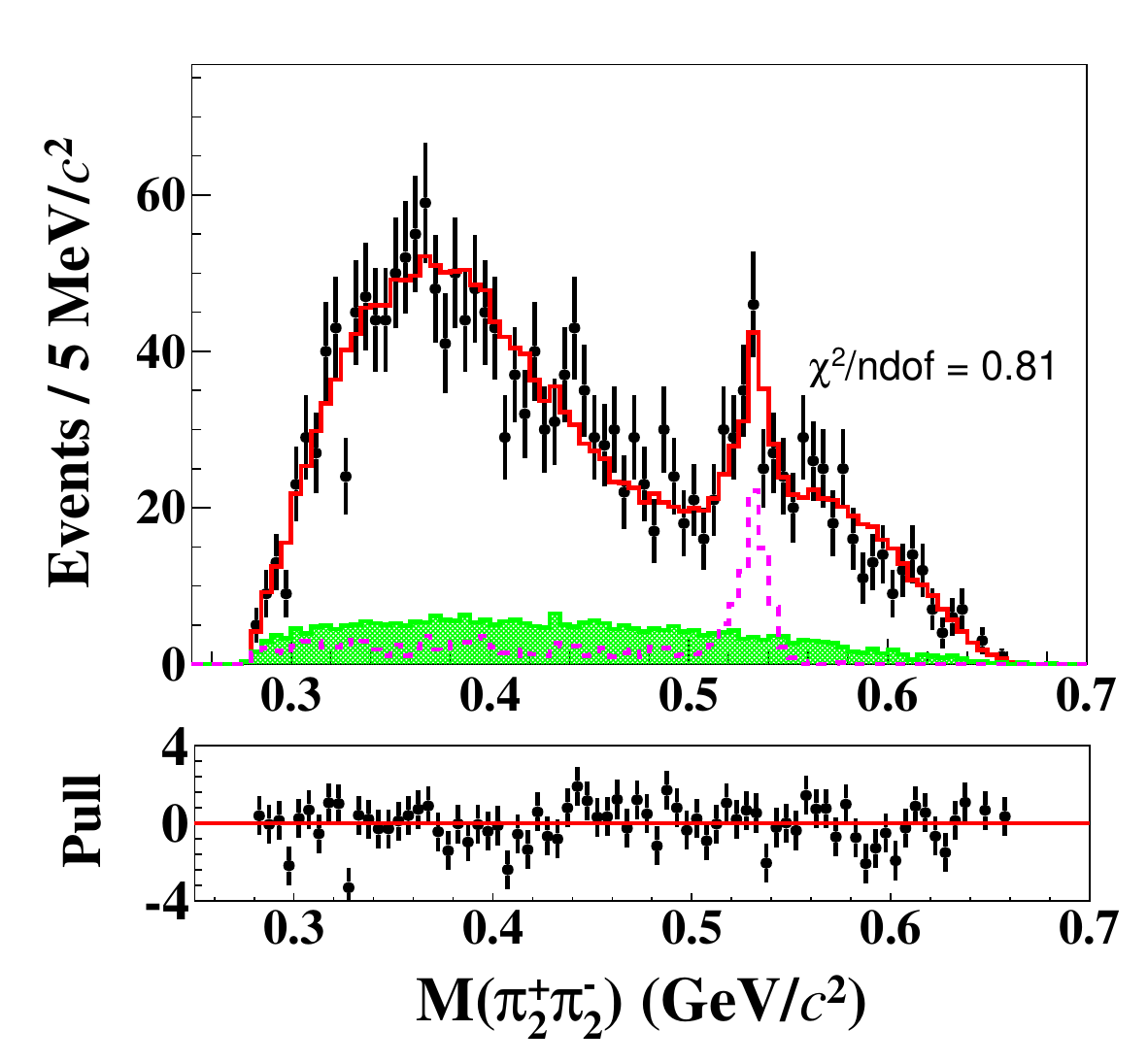}\put(-30,99){\bf~(d)}
\caption{The fitted projections to the invariant mass of four $\pi^{+}\pi^{-}$  combinations, where $\pi^{+(-)}_{1}$ and $\pi^{+(-)}_{2}$ are sorted in ascending order of their energy.
}
\label{fig:tff_fig}
\end{center}
\end{figure*}

To simplify the fit,   the mass and width of the  $\rho$ are fixed at the world-average values~\cite{pdg}, $M_\rho=775.11$~MeV and $\Gamma_\rho=149.1$~MeV.
The amplitude analysis fit gives $\alpha =1.22 \pm 0.33$, where the uncertainty is statistical only. Projections of the data and fit results to the mass spectra of different $\pi^+\pi^-$ combinations are displayed in Fig.~\ref{fig:tff_fig}, which indicates that the amplitude fit provides a good description of the data.
An alternative fit with free resonant parameters of $\rho$ is also performed, which returns a fitted mass   $M_\rho=671.9\pm13.4$~MeV and width $\Gamma_\rho=233.7\pm16.7$~MeV that deviate significantly from the world-average values. In this case, we also find that the fit result for $\alpha$ is not stable because the terms with and without $\alpha$ in Eq.~\ref{eqfit_amp} have a similar lineshape, which leads to a strong correlation between them. 
Thus, the fit with the fixed $\rho$ parameters is taken as the baseline. This measurement is in good agreement with the prediction of unity from the combined ChPT and VMS models in Ref.~\cite{GuoFK}, where $c_3=1$ and $c_1 -c_2=1$.

\section{Search for $\eta' \to 4 \pi^{0} $}
\begin{figure}[htp]
  \begin{center}
    \subfigure{
      \label{m4pi0}
      \includegraphics[width=0.49\textwidth]{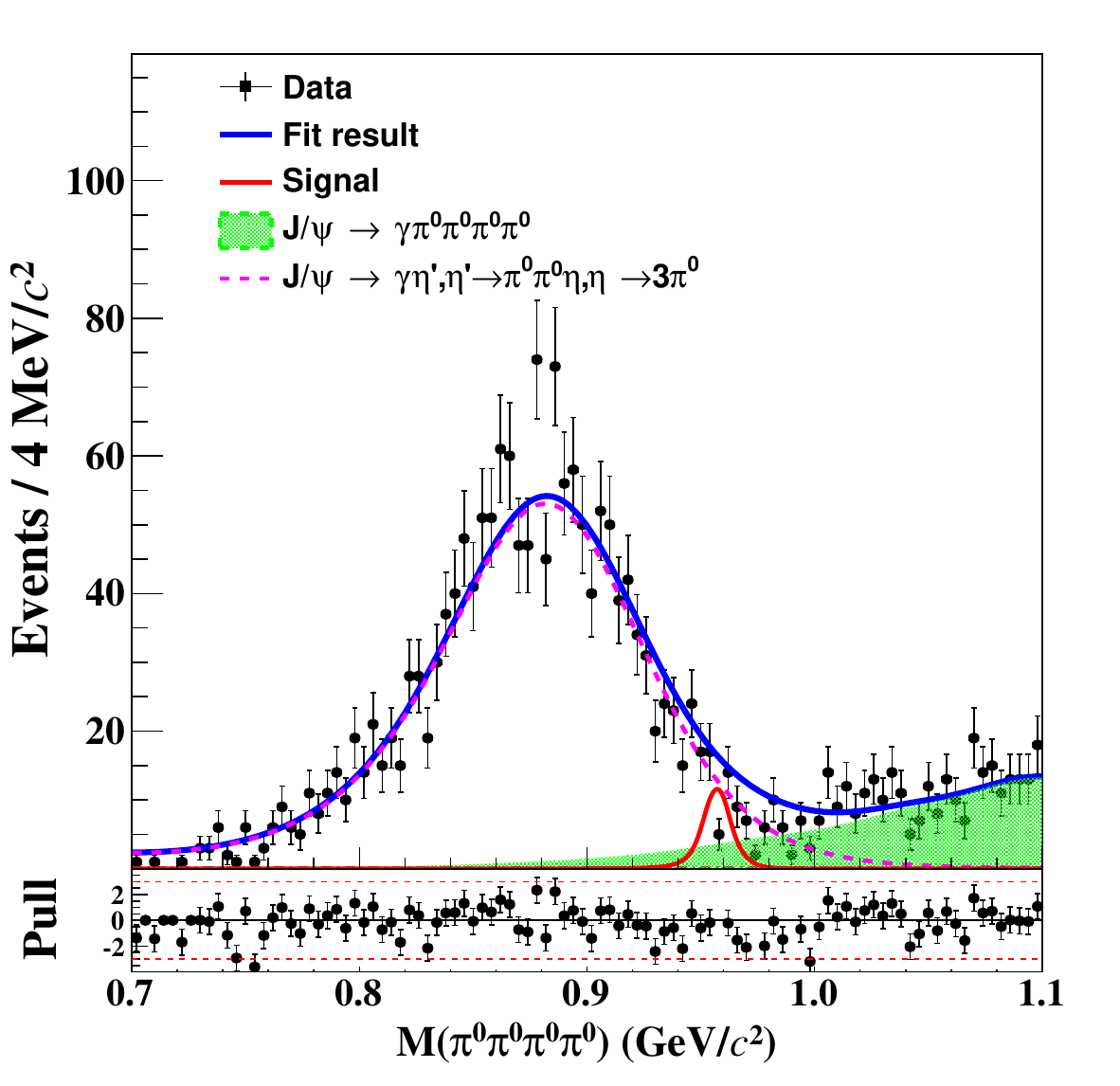}
      \put(-40,215){\bf~(a)}
    }
    \subfigure{
      \label{likelihood}
      \includegraphics[width=0.49\textwidth]{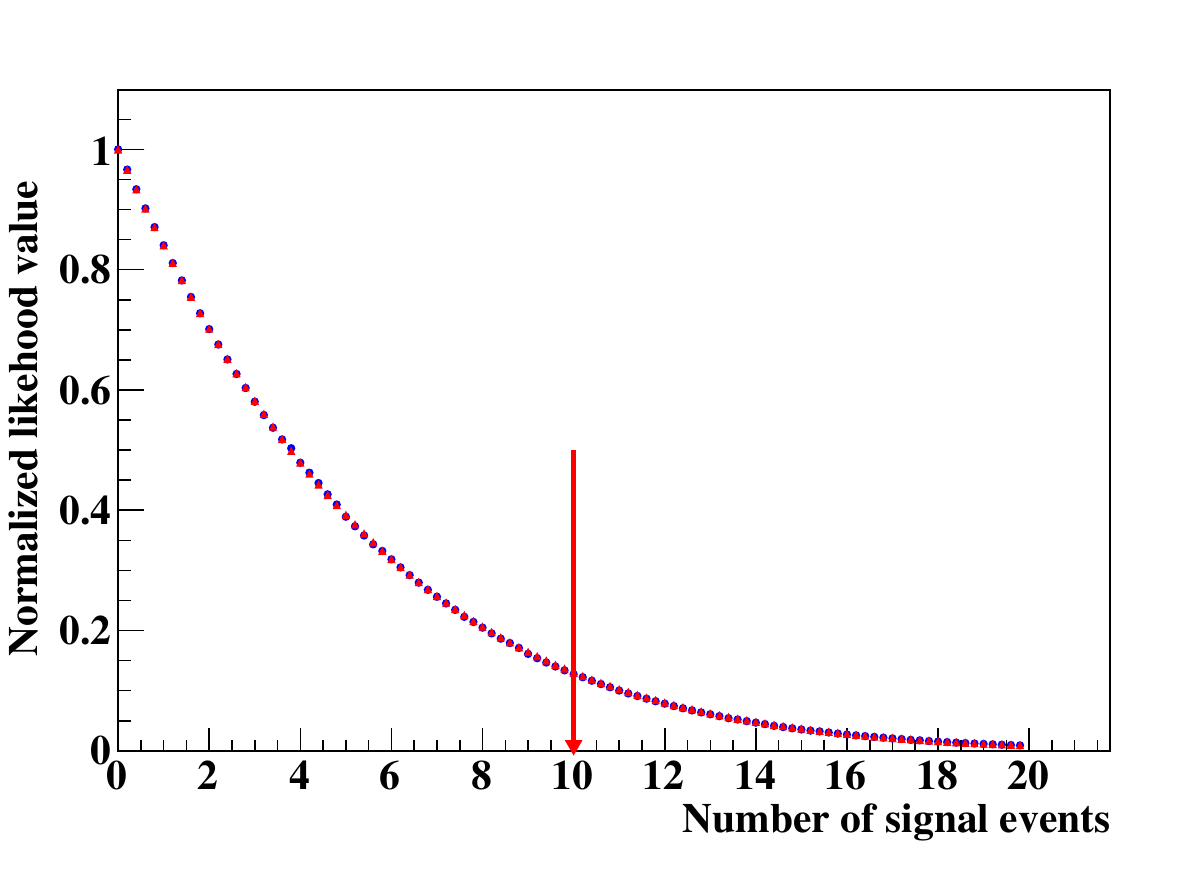}
      \put(-40,155){\bf~(b)}
    }
    \caption{The fit result to the 4$\pi^{0}$ invariant mass
(a) and the normalized likelihood distribution (b). 
In (b), the blue dots and red triangles are the likelihood distributions before and after convolution with a Gaussian function with resolution equal to the total systematic uncertainty, respectively.  The red arrow indicates the upper limit at the 90\% confidence level.}
    \label{fig:m4pi0_mass}
  \end{center}
\end{figure}

No significant signal contribution is evident for  the decay $\eta' \to 4 \pi^{0}$ in the $M(4\pi^{0})$ distribution shown in Fig.~\ref{m4pi0}.
To quantify this conclusion,  an unbinned maximum-likelihood fit is performed on the 4$\pi^{0}$ mass spectrum, allowing for background contributions and a possible signal component.
The total PDF consists of a signal component,  the peaking and non-peaking background contributions.
The signal component is modeled as the MC simulated shape. Both the peaking and
non-peaking background contributions from $J/{\psi} \to \gamma \eta', \eta' \to \pi^{0} \pi^{0} \eta, \eta \to \pi^{0} \pi^{0} \pi^{0}$ and $J/{\psi} \to \gamma \pi^{0} \pi^{0}  \pi^{0} \pi^{0}$ are also described with shapes obtained from
the dedicated MC simulation.

A Bayesian approach \cite{bayesian} is used to determine an upper limit on the branching fraction of $\eta' \to 4 \pi^0$.  A series of fits are performed for different assumed values $N$ of the signal yield, and
the negative log-likelihood $\mathcal{S}$ is determined for each fit.
With the upper limit of $N_{\rm UL}$, the upper limit on the branching fraction is calculated with 
\begin{equation}\label{eqBr0_BF2_1}
\Br\left(  {\eta^\prime}\to 4\pi^{0}\right) < {\textstyle{N_{\rm UL} \over {{N_{J/{\psi}}} \cdot  {\Br}\left( {{J/{\psi}} \to \gamma \eta' } \right) \cdot {\Br^4}\left( {{\pi^{0}} \to \gamma \gamma } \right) \cdot \epsilon }}},
\end{equation}
where $N_{J/{\psi}}$ is the total number of $J/{\psi}$ events, $\epsilon$ is the detection efficiency obtained by MC simulation, $\mathcal {B}(J/{\psi}\to\gamma \eta')$ and $\mathcal {B}({\pi^{0}}\to\gamma \gamma)$  are quoted from the PDG~\cite{pdg}, respectively.

The distribution of normalized likelihood values, defined as $\mathcal{L}(N)=\rm exp(-[\mathcal{S}(N) - {\mathcal{S}}_{\rm min}])$, where ${\mathcal{S}}_{\rm min}$ is the minimum negative log-likelihood obtained from the ensemble of fits, is taken as the PDF for the likelihood function $\mathcal{L}({N})$.
The upper limit on the number of signal events, ${N}_{\rm UL}$, is determined at 90$\%$ of the integral of the PDF
\begin{equation}\label{eqBr0_BF2_2}
 {\textstyle{{\int}^{{N}_{\rm UL}}_0 {\mathcal{L}}({N}) d{N} \over {\int}^{\infty}_0 {\mathcal{L}}({N}) d {N}}} = 0.9,
\end{equation}
where $N$ is the fitted number of signal events. To account for the additive systematic uncertainties related to the fits, several alternative fits are performed. These alternative fit involve different fit ranges, continuum background shapes and peaking background shapes. Among the results of these fits, the largest number of signal events $9.9$ is chosen to calculate the upper-limit for the branching fraction at the $90\%$ confidence level, resulting in ${\mathcal{B}}_{\rm UL}=1.22 \times 10^{-5}$.

\section{Systematic uncertainty}\label{sys}
Sources of systematic uncertainties and their corresponding contributions to the measurements of the branching fractions for the different decay modes are summarized in Table~\ref{list_sys}.
The total systematic uncertainties are obtained by adding all the contributions in quadrature under the assumption that they are independent.

The MDC tracking efficiency is
studied using a clean control sample of $J/\psi \to \pi^{+} \pi^{-} \pi^{0}$ decays. It is found that the MC simulation agrees
with data within 0.5\% for each charged track. The systematic
uncertainties on the tracking efficiency
are taken as 2\% and 1\%, for Mode I and Mode II, respectively. 

The photon-detection efficiency
is studied with a dedicated sample of the $e^{+} e^{-} \to \gamma \mu^{+} \mu^{-}$ events. 
The difference in efficiencies between data and MC simulation, 0.5\%, is taken as the corresponding systematic uncertainty.

Uncertainties associated with the kinematic fits for the charged decay Mode I and Mode II come from
the inconsistency of the track helix parameters between
data and MC simulation. The helix parameters for the
charged tracks of the MC samples are corrected to eliminate this 
inconsistency, as described in Ref.~\cite{4Ckfit}, and the agreement
of $\chi^{2}$ distributions between data and MC simulation is
much improved. We take half of the differences on the
selection efficiencies with and without the correction as the
systematic uncertainties, which are $0.6\%$ and $0.2\%$ for Mode I and Mode II, respectively.
The uncertainty for Mode III associated with the kinematic fit is estimated by adjusting the components of the photon-energy error matrix in the signal MC sample to reflect the known difference in resolution between data and MC simulation \cite{8Ckfit1} and the difference is estimated to be $5.6\%$.

Uncertainties arise from the understanding of the background associated with the estimation of the resonant and non-resonant background events.  For the background contributions from the phase-space events, e.g., $J/\psi\rightarrow\gamma\pi^+\pi^-\pi^+\pi^-$
and $J/\psi\rightarrow\gamma\pi^+\pi^-\pi^0\pi^0$, alternative fits are performed by replacing phase-space MC shapes with a third-order Chebychev function.  The deviations in signal
yields after implementing these modifications are found to be $0.2\%$ and $1.6\%$ for Mode I and Mode II, respectively. 
 The uncertainties due to peaking background events below or just above the $\eta'$ peak are considered by fixing these contributions according to the branching
fractions of $J/{\psi} \to \gamma \eta'$ and the cascade decays, and lead to uncertainties of $0.4\%$ and $0.1\%$, respectively.

The uncertainty associated with the $\eta$ and $\omega$ veto is studied by varying the $\eta$  and $\omega$ mass requirements, and is assigned to be $2.8\%$ for mode II.
The uncertainty from the MC model is estimated by varying the parameters in the ChPT and VMD model,
and the largest difference in the detection efficiencies is assigned as the systematic uncertainty, which is  $0.8\%$ for mode I, $0.6\%$ for Mode II and $0.1\%$ for Mode III.
 Similarly, the systematic uncertainty associated with vetoing $\eta$ signal in Mode III is estimated to be $3.0\%$. The uncertainties in the 
branching fractions of $J/{\psi}\to\gamma \eta'$ and $\pi^{0}\to\gamma \gamma$ are $1.3\%$ and $0.06\%$, respectively.
The uncertainty due to the total number of $J/{\psi}$ events ($N_{J/{\psi}}$) is $0.44\%$ \cite{yanghx}.

\begin {table}[htbp]
\begin{center}
\begin{small}
    {\caption {The systematic uncertainties on the branching-fraction measurements (in units of $\%$).}
    \label{list_sys}}
    \begin {tabular}{l c c c}\hline\hline
	Source & ~~Mode I~~ &  ~~Mode II~~& ~~Mode III~~                        \\   \hline	
	MDC tracking  &     $2$      &  $1$                              & $-$      \\
 	Photon detection    &     $0.5$      &  $2.5$                    & $4.5$                   \\
	Kinematic fit   &     $0.6$     &  $0.2$                         & $5.6$                 \\
	Peaking background &    $0.4$      &  $0.1$                     & $-$                    \\
    Continuous background   &     $0.2$     & $1.6$ 	             & $-$                  \\
	Veto $\eta(\omega)$ signal      &     $-$     &  $2.8$           & $-$                    \\
    Veto $\eta \to 3 \pi^{0}$       &     $-$     &  $-$             & $3.0$                  \\
    $\mathcal {B}(J/{\psi}\to\gamma \eta')$	&  $1.3$     &  $1.3$    & $1.3$                   \\
    $\mathcal {B}(\pi^{0}\to\gamma \gamma)$	&  $-$      &  $0.06$    & $0.12$                  \\
    $N_{J/\psi}$ &  $0.44$       & $0.44$             & $0.44$                  \\
    Generator model  &  0.8       & 0.6                              & $0.1$    \\             \hline
	 Total & 2.7 & 4.5  & $8.0$       \\

\hline
\hline
\end{tabular}
\end{small}
\end{center}
\end{table}

The systematic uncertainties on the doubly virtual isovector contribution $\alpha$ from the amplitude analysis of $\eta^\prime\to\pi^+\pi^-\pi^+\pi^-$ are summarized in Table \ref{list_sys3}.
The uncertainty in the pion tracking efficiency is studied using a control sample of $J/{\psi} \to \pi^{+} \pi^{-} \pi^{0}$ decays. The data-MC difference  is extracted as a function of the particle momentum and the cosine of the polar angle, and then a reweighting procedure is applied on the charged pions for each event to calculate the MC integral cross section. Following this, an alternative amplitude analysis is conducted and the difference between the results with and without correction,  $0.1\%$, is assigned as the systematic uncertainty.
 For  the continuum background  from  $J/\psi \rightarrow \gamma \pi^{+} \pi^{-} \pi^{+} \pi^{-}$,  by replacing with a third-order Chebychev polynomial function in the fit, the normalized background events is found to be  $273.1 \pm 16.5$. 
An alternative amplitude fit is performed and the change in $\alpha$  with respect to the nominal result, 2.0\%, is taken as the systematic uncertainty. 
The uncertainty from peaking background associated with the background events ($J/{\psi} \to \gamma \eta', \eta' \to \pi^{+} \pi^{-}\eta, \eta \to \gamma \pi^{+} \pi^{-}$) is estimated by
normalizing the number of background events  to be $177.6 \pm 13.3$ in the amplitude analysis in accordance with  the branching fractions of $J/{\psi} \to \gamma \eta'$ and the cascade decays~\cite{pdg}. 
The change of result, $2.9\%$,  is taken as the systematic uncertainty.
The systematic uncertainties associated with the mass and width of $\rho$, are estimated by varying the range by $\pm 1\sigma$ instead of fixing to the world average values, which are determined to be 0.7\% and 0.2\%, respectively.
The total systematic uncertainty of 3.6\% is obtained by adding all systematic uncertainties in quadrature assuming that they are independent.

\begin {table}[htbp]
\begin{center}
\begin{small}
    {\caption {The systematic uncertainties on the doubly virtual isovector contribution $\alpha$ to $\eta' \to \pi^{+} \pi^{-} \pi^{+} \pi^{-}$ (in unit of $\%$).}
    \label{list_sys3}}
    \begin {tabular}{l c c  c}\hline\hline
	Source & $\eta' \to \pi^{+} \pi^{-} \pi^{+} \pi^{-}$           \\   \hline
 	MDC tracking             &     $0.1$                                       \\
    4C kinematic fit         &   $0.5$                                 \\
    Continuum background    &     $2.0$                          \\
    Peaking background       &     $2.9$                          \\
	Mass of $\rho$       &     $0.7$                          \\
	Width of $\rho$      &     $0.2$                          \\\hline
	 Total &      $3.6$    \\
\hline
\hline
\end{tabular}
\end{small}

\end{center}
\end{table}

\section{Branching-fraction results}

After taking into account the systematic uncertainties, the branching fractions of $\eta' \to \pi^{+} \pi^{-} \pi^{+} \pi^{-}$ and $\eta' \to \pi^{+} \pi^{-} \pi^{0} \pi^{0}$ are measured to be
$\left( {8.56 \pm 0.25({\rm stat.}) \pm 0.23({\rm syst.})} \right) \times {10^{ - 5}}$
and $\left( {2.12 \pm 0.12({\rm stat.}) \pm 0.10({\rm syst.})} \right) \times {10^{ - 4}}$, respectively, where the first uncertainties are statistical and the second systematic.
No $\eta' \to 4\pi^{0}$ signal is observed and the upper limit on its decay branching fraction, considering only additive systematic uncertainties is found to be ${\mathcal{B}}_{UL}=1.22 \times 10^{-5}$
at the $90\%$ confidence level.  The final upper limit on the branching fraction also considers the multiplicative systematic uncertainty, which is accounted for 
 by  convolving the likelihood distribution $L(N)$  to obtain the smeared likelihood $L'(N)$,
\begin{equation}\label{eqBr0_BF2_3}
  L'(N)={\displaystyle{{\int}^{1}_{0}} \textstyle{ L  ( \frac{S}{\hat{S}}  N)  {\rm exp}  \left[ - \frac{\left(S - \hat{S}\right)}{2 {\sigma}^2_{S}}  \right]  d S  }}.
\end{equation}
In this expression $\hat{S}$ is the nominal efficiency,
$\sigma_{S}$ is its systematical uncertainty ($8\%$) coming from Table \ref{list_sys} and $N$ is the number of signal events for $\eta' \to 4\pi^{0}$.
Following this procedure, the upper limit on the number of signal events at the $90\%$ confident level is set to be $10$ as shown in Fig.~\ref{likelihood}, 
and the corresponding upper limit of the branching fraction is $\mathcal{B}(\eta' \to 4 \pi^{0})<1.24 \times {10^{ - 5}}$.

\section{Summary}
In summary,  using a  sample of $(10087 \pm 44) \times {10^{6}}$ $J/{\psi}$ events, we perform improved measurements of the branching fractions of  $\eta^\prime\to \pi^{+} \pi^{-} \pi^{+} \pi^{-}$ and $\eta^\prime\to\pi^{+} \pi^{-} \pi^{0} \pi^{0}$ decays with $J/\psi\rightarrow\gamma\eta^\prime$. The measured branching fractions of these two decays  are in good agreement with the previous works~\cite{A1-bes3-eta4pi-lihj}, with a three-times improvement in precision. In addition, within the framework of the ChPT and VMD model, an amplitude analysis of  $\eta^\prime\to \pi^{+} \pi^{-} \pi^{+} \pi^{-}$ is performed for the first time to
investigate the doubly virtual isovector contribution. The parameter $\alpha$, corresponding to $\frac{c_1-c_2}{c_3}$ in Ref.~\cite{GuoFK}, is determined to be  $1.22 \pm 0.33({\rm stat.}) \pm 0.04({\rm syst.})$, which is consistent with the prediction  of  theoretical calculations~\cite{GuoFK}.

 We have also searched for the rare decay  $\eta' \to 4 \pi^{0}$, and no evidence is found. The upper limit of $\mathcal{B}(\eta' \to 4 \pi^{0}) $ at the 90\% confidence level is determined to be $1.24 \times 10^{-5}$, thereby improving on the previous limit~\cite{A2-bes3-eta4pi-chang} by a factor of four.

\section{Acknowledgement}
The BESIII Collaboration thanks the staff of BEPCII and the IHEP computing center for their strong support. This work is supported in part by National Key R\&D Program of China under Contracts Nos. 2020YFA0406300, 2020YFA0406400; National Natural Science Foundation of China (NSFC) under Contracts Nos. 11635010, 11735014, 11835012, 11935015, 11935016, 11935018, 11961141012, 12025502, 12035009, 12035013, 12061131003, 12192260, 12192261, 12192262, 12192263, 12192264, 12192265, 12221005, 12225509, 12235017; the Chinese Academy of Sciences (CAS) Large-Scale Scientific Facility Program; the CAS Center for Excellence in Particle Physics (CCEPP); Joint Large-Scale Scientific Facility Funds of the NSFC and CAS under Contract No. U1832207; CAS Key Research Program of Frontier Sciences under Contracts Nos. QYZDJ-SSW-SLH003, QYZDJ-SSW-SLH040; 100 Talents Program of CAS; The Institute of Nuclear and Particle Physics (INPAC) and Shanghai Key Laboratory for Particle Physics and Cosmology; European Union's Horizon 2020 research and innovation programme under Marie Sklodowska-Curie grant agreement under Contract No. 894790; German Research Foundation DFG under Contracts Nos. 455635585, Collaborative Research Center CRC 1044, FOR5327, GRK 2149; Istituto Nazionale di Fisica Nucleare, Italy; Ministry of Development of Turkey under Contract No. DPT2006K-120470; National Research Foundation of Korea under Contract No. NRF-2022R1A2C1092335; National Science and Technology fund of Mongolia; National Science Research and Innovation Fund (NSRF) via the Program Management Unit for Human Resources \& Institutional Development, Research and Innovation of Thailand under Contract No. B16F640076; Polish National Science Centre under Contract No. 2019/35/O/ST2/02907; The Swedish Research Council; U. S. Department of Energy under Contract No. DE-FG02-05ER41374.

  
\end{document}